\documentstyle[prl,aps,multicol,epsf,times]{revtex}

\begin{document}
\draft 
\title{Dynamical simulation of current fluctuations in a dissipative
two-state system}

\author{J\"urgen T. Stockburger and C.H. Mak}
\address{Department of Chemistry, University of Southern California,
Los Angeles, CA 90089-0482, USA}

\date{Received October 23, 1997}
\maketitle
\begin{abstract}
Current fluctuations in a dissipative two-state system have been
studied using a novel quantum dynamics simulation method.  After a
transformation of the path integrals, the tunneling dynamics is
computed by deterministic integration over the real-time paths under
the influence of colored noise. The nature of the transition
from coherent to incoherent dynamics at low temperatures is
re-examined.
\end{abstract}


\begin{multicols}{2} 
\narrowtext A two-state system coupled to a dissipative environment is
the archetypical model for tunneling phenomena in condensed phase.  It
has found widespread applications in solid-state physics
\cite{gw,glass,Chakravarty-Kivelson}, most recently in interlayer
charge transport in high-$T_c$ superconductors
\cite{Chakravarty-Anderson} and quantum computing\cite{garg}, as well
as in biophysics \cite{marcus} for the modeling of electron transport
in biochemical reactions.  One of the most intriguing features of this
model is a dynamical phase transition between coherent tunneling and
incoherent relaxation. This was first predicted by Chakravarty and
Leggett \cite{Chakravarty-Leggett,Leggett} and later confirmed by
experiments on interstitial tunneling in niobium\cite{Wipf}.

Although the existence of the coherent-incoherent transition is widely
accepted, its precise nature and location has been called
into question by some recent calculations \cite{Lesage,Costi,Egger,Strong}.
Coherence is a phenomenon of dynamics, yet an exact
treatment of tunneling in the time domain has so far been out of reach.
The original prediction of the transition \cite{Leggett} was based on
a dynamical but approximate theory, whereas the more recent theories,
suggesting the transition would occur at a much weaker damping than
predicted earlier, were based on 
statistical mechanical calculations \cite{Lesage,Costi}.

In this Letter, we describe a new exact numerical method for calculating the
real-time dynamics of dissipative quantum systems and use it to
investigate the transition from coherent to incoherent dynamics in a
two-state system at low temperatures.
Previously, the only exact numerical approach to tunneling dynamics has 
been the dynamical quantum Monte Carlo (QMC) method \cite{Egger,Leung}.
But all real-time QMC 
simulations fail at longer times because the signal-to-noise ratio of
the results vanishes exponentially due to the highly
oscillatory integrand. This problem is commonly
referred to as the {\em dynamical sign problem}. In the case of large
bandwidth of the dissipative environment, QMC simulations further suffer
from a {\em slowing-down problem} caused by the increasingly long-lived
correlations in the sampling process. 
The new method eliminates both problems through a generalized 
Hubbard-Stratonovich transformation and 
allows us to perform the functional integration over paths of
the tunneling system by a deterministic method, while
statistically sampling fluctuations from an ensemble of Gaussian noise
trajectories.

Dissipative two-state systems are often described by the spin-boson
model,
\begin{eqnarray} \label{HSB}
H &=& -{1\over 2} \Delta \sigma_x + \sum_\nu \omega_\nu (a^\dagger_\nu a_\nu
+ {1\over 2} )\nonumber\\
&& + \hat{q} \sum_\nu C_\nu (a_\nu + a^\dagger_\nu ),
\end{eqnarray}
where $\hat{q} = q_0 \sigma_z\,/2$ is the position operator of the
tunneling system with intrinsic tunneling frequency $\Delta$,
$\sigma_x$ and $\sigma_z$ are Pauli spin matrices, and $\hbar = 1$.
The effect of the harmonic environment is fully characterized by a
spectral density $ J(\omega) = \pi \sum_\nu C_\nu^2 \,
\delta(\omega-\omega_\nu ) $, for which the Ohmic form $ J(\omega) =
2\pi \alpha \omega / q_0^2$ is experimentally the most relevant and
theoretically the most interesting.  The Ohmic spectral density
introduces a single dimensionless damping constant $\alpha$. This
model must be regularized by an upper cutoff $\omega_{\rm c}$ of the
spectral density.  The scaling limit $\omega_{\rm c} \gg \Delta$ is
characteristic of tunneling in solids and as shown by scaling
arguments\cite{Chakravarty}, the Ohmic spin-boson model has nontrivial
dynamics only for $\alpha < 1$, and the renormalized tunneling
frequency
\begin{equation} \label{Delta_r}
\Delta_{\rm r} = \Delta \left(\Delta/\omega_{\rm c}\right)^{\alpha/(
1-\alpha)},
\end{equation}
is the only frequency scale of the dynamics at zero temperature other
than $\omega_{\rm c}$.  The transition from coherent to incoherent
dynamics occurs at a critical damping $\alpha_{\rm c}$ at which the
$Q$ factor of the tunneling oscillations vanishes. Tunneling
oscillations can be observed as a damped oscillatory component of the
position correlation function $C(t) = {1\over 2}\, \langle
\sigma_z(t)\sigma_z(0)+\sigma_z(0)\sigma_z(t)\rangle$ that is present
in addition to an incoherent relaxation background\cite{Guinea}. The
asymptotic long-time behavior is always dominated by an algebraic
incoherent decay, $C(t) \propto \alpha \Delta_{\rm r}^{-2}
t^{-2}$\cite{alge-remark}.

A coherence criterion equivalent to finite $Q$ is a finite {\em
dephasing time} of the quantum beats that manifest themselves as
tunneling oscillations. A measure of this dephasing time is given by
the lifetime of delocalized states of the tunneling system
\cite{Zurek} (see also\cite{Strong,Stuttgart}) $|+\rangle \pm
i|-\rangle$, which are eigenstates of the tunneling current $j =
\Delta\sigma_y/2 = \dot\sigma_z/2$. The correlation time $\tau$ of the
current correlation function
\begin{equation}
C_{jj}(t) = {1\over 2}\, \langle j(t)j(0)+j(0)j(t)\rangle
\end{equation}
equals the lifetime of these superposition states. A finite
correlation time of the {\em current}\/ correlation function thus
implies coherent oscillations in the {\em position}\/ correlation
function. This relationship allows us to identify coherence even for
very strongly damped cases in which oscillations may be masked by the
incoherent background. For a two-state system, the relation
$\ddot C(t) = -4\, C_{jj}(t)$
provides another direct connection to previous studies on the position
correlation function.

To compute the exact dynamics of $C_{jj}$, we employ a hybrid
stochastic/deterministic numerical method which we shall label
chromostochastic quantum dynamics (CSQD).  This method, which is
generally applicable to quantum systems with linear dissipation, is
based on the path integral formulation of dissipative quantum
dynamics\cite{Leggett,WeissBuch}.  The time evolution of the reduced
density matrix for the system coordinate $q$ can formally be
represented by a double functional integral\cite{Feynman-Vernon}
\begin{equation} \label{rho_PI}
\rho(q_{\rm f},q'_{\rm f};t) =
\int_{q_{\rm i}}^{q_{\rm f}} \!\!\!{\cal D}[q]
\int_{q'_{\rm i}}^{q'_{\rm f}} \!\!\!{\cal D}[q']
e^{i S_0[q] - i S_0[q']} F[q,q'].
\end{equation}
$S_0[q]$ is the action of the undamped quantum system,
and its interactions with the environment are  
incorporated into a complex-valued influence functional $F[q,q'] =
\exp(-\Phi' - i\Phi'')$ with 
\begin{eqnarray}
\Phi'[q,q'] &=& \int_{t_0}^t \!\!\!dt' 
\int_{t_0}^{t'} \!\!\!dt''
(q(t')-q'(t')) \nonumber\\
&& \times L'(t'-t'') (q(t'')-q'(t'')), \\ \label{phipp}
\Phi''[q,q'] &=& {1\over 2} \int_{t_0}^t \!\!\!dt'
\int_{t_0}^{t'} \!\!\!dt''
(q(t')-q'(t')) \nonumber\\
&& \times m\gamma(t'-t'') (\dot q(t'')+\dot q'(t'')).
\end{eqnarray}
$L'(t)$ is the real part of the autocorrelation function of the
collective bath mode $\sum_\nu C_\nu (a_\nu + a^\dagger_\nu )$, 
averaged over an ensemble of free oscillators, $m$
is the mass of the tunneling particle, and $\gamma(t)$ is the
classical friction kernel associated with the spectral density
$J(\omega)$.
This influence functional describes a `factorized' initial
preparation, i.e. with the particle constrained to one side for times 
$t<t_0$ with the environment fully relaxed.
To simulate equilibrium correlation functions, it is necessary to 
push the preparation back to a sufficiently large negative time $t_0<0$ 
\cite{WeissBuch} and insert measurement operators $j$ at times $0$ and $t$.

A direct quantum Monte Carlo evaluation of a time-discretized
version of type (\ref{rho_PI}) can be prohibitively expensive due to
the dynamical sign and slowing-down problems. An
alternative approach for single-particle problems or
dissipative systems near the Markovian limit is the explicit
iteration of a short-time propagator. This reduces the path integral
to a manageable series of matrix multiplications\cite{Thirumalai}.
Another algorithm was recently presented by Cao, Unger and Voth\cite{Cao} 
for an environment with a few oscillators. By directly sampling the oscillator
paths and propagating the system coordinate for each sample, they
eliminate memory effects.

In this Letter, we tackle the problem of memory effects in the important case of
a {\em continuum} of environmental modes, such as in the case of Ohmic 
friction, where sampling over individual oscillator trajectories is not 
feasible. The major obstacle in trying
to decompose the path integral (\ref{rho_PI}) into short-time
propagators lies in the interaction kernel
$L'(t)$ because its range 
diverges as the temperature approaches zero. In comparison,
the friction kernel $\gamma(t)$ poses no problem because it
vanishes at time larger than $\omega_{\rm c}^{-1}$, the
shortest timescale in our problem.

The problem with the long-range interactions introduced by $L'(t)$ can 
nonetheless be solved, albeit at the cost of introducing an additional 
path variable.  The exponential of the non-local action 
$\Phi'[q,q']$ can be decomposed into a superposition of time-local 
phase factors,
\begin{eqnarray} \label{noise}
&&\exp(-\Phi'[q,q']) = \nonumber \\
&&\int {\cal D}[\xi] \, W[\xi] 
\exp\left\{ -i 
\int_{t_0}^t \!\!\!dt' \xi(t')(q(t')-q'(t')) \right\}.
\end{eqnarray}
The distribution $W[\xi]$ is real and Gaussian, with
$\langle\xi(t)\xi(t')\rangle_W = L'(t-t')$, and normalizable
through the condition $\Phi'[q,q] \equiv 0$. Formally, this
decomposition is a Hubbard-Stratonovich transformation in a function
space over the interval $[t_0,t]$. Equation (\ref{noise}) is equivalent
to the construction of an influence functional for a
classical colored noise source\cite{Feynman-Vernon}, and as such, we
will interpret the function $\xi(t)$ as a noise trajectory.

In the CSQD algorithm, the propagation of the system coordinates $q$
and $q'$ is
carried out deterministically for each realization of the noise
trajectory, while the noise trajectories are sampled
%stochastically
from the distribution $W[\xi]$.
Instead of generating weights from a Metropolis-type random walk, 
statistical weights are assigned to noise trajectories {\em a priori} by
numerically filtering white noise.

In general, there are two ways to treat the remaining term $\Phi''$.
If $\hat q$ acts on a finite-dimensional Hilbert space, $q(t)$
contains discrete jumps between the eigenvalues of $\hat q$.  The
`state vector' that is propagated must then remember these `virtual
transitions' during a finite number of preceding time slices. But
since the friction kernel $\gamma(t)$ decays rapidly over the memory
time $\omega_{\rm c}^{-1}$, the number of transitions needed `in
memory' is small for large cutoff. On the other hand, if $\hat q$
describes a continuous degree of freedom, the kinetic term in
$S_0[q]$ requires the relevant paths $q(t)$ to be smooth. Then the
limit $\omega_{\rm c} \to \infty$ can be applied to (\ref{phipp}),
making $\Phi''[q,q']$ a time-local functional.

For the two-state system, we achieve excellent accuracy with a maximum
of two tunneling events `in memory'. The error resulting from this
truncation can be made arbitrarily small by increasing $\omega_{\rm
c}$, i.e., moving further into the scaling
regime\cite{cutoff-remark}. The length of our time steps is a fixed
fraction of the inverse bandwidth $\omega_c^{-1}$, which is short
enough to make the particular choice of approximate short-time
propagator a minor issue.

Empirically, we find that the statistical variance of CSQD grows no
faster than {\em logarithmically} with $t$.  This remarkable
performance compared to the {\em exponential} divergence observed in
conventional QMC methods results from the fact that deterministic
integration is not affected by the oscillatory nature of the
integrand.  Consecutive samples in the CSQD simulation are by design
{\em statistically independent}. This completely eliminates any
slowing-down problem. We have tested the validity and accuracy of our
method, comparing to known analytic results for the real-time
fluctuations and the relaxation behavior of the tunneling coordinate
at damping $\alpha=1/2$ and $\alpha\ll 1$. We find excellent agreement
between our numerical results and previous analytic calculations.

\begin{figure}
\epsfxsize=0.8\columnwidth
\centerline{\epsffile{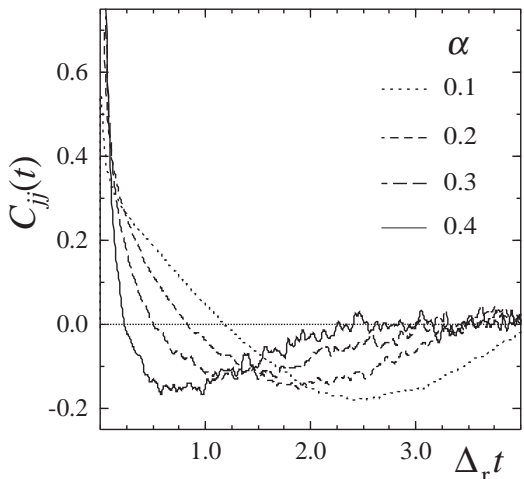}}
\caption[]{Current correlation function for various damping\linebreak
$\alpha<1/2$, $\omega_{\rm c}/\Delta_{\rm r} = 50$.
\label{Fig-mixed}}
\end{figure}

\begin{figure}
\epsfxsize=0.8\columnwidth
\centerline{\epsffile{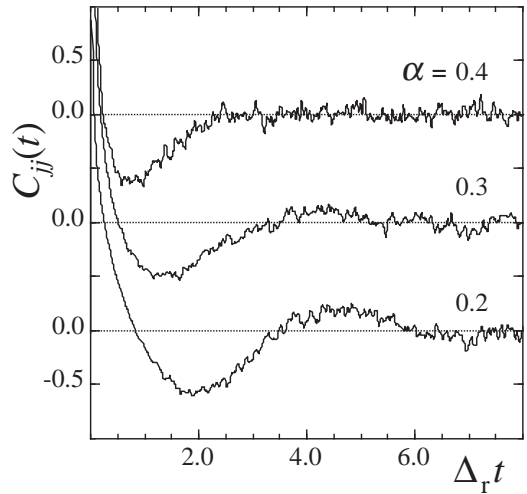}}
\caption[]{Coherent oscillations and long-time decay of the current
correlation function ($\omega_{\rm c}/\Delta_{\rm r} =
50$).\label{Fig-oscill}}
\end{figure}

Fig.\ \ref{Fig-mixed} shows CSQD results for $C_{jj}(t)$ at zero
temperature and for $0<\alpha<1/2$. Except for the initial drop at times
$t\lesssim \omega_{\rm c}^{-1}$, all four curves represent functions
of the scaling argument $\Delta_{\rm r}t$ only.  To obtain a formal
estimate for $\tau$, we define it to be the first zero of $C_{jj}(t)$
\cite{deph-remark}.  Positive current-current correlations do persist
up to times of order $\Delta_{\rm r}^{-1}$, i.e., the scaled dephasing
time $\Delta_{\rm r}\tau$ is nonzero and finite (at least) up to
$\alpha=0.4$, although it declines significantly with increasing
$\alpha$.
Quantum coherence becomes increasingly short-lived, however, making it
impossible to resolve oscillations from a background of monotonous
signal decay for $\alpha\gtrsim 0.3$ (Fig.\ \ref{Fig-oscill}).

\begin{figure}
\epsfxsize=0.95\columnwidth
\centerline{\epsffile{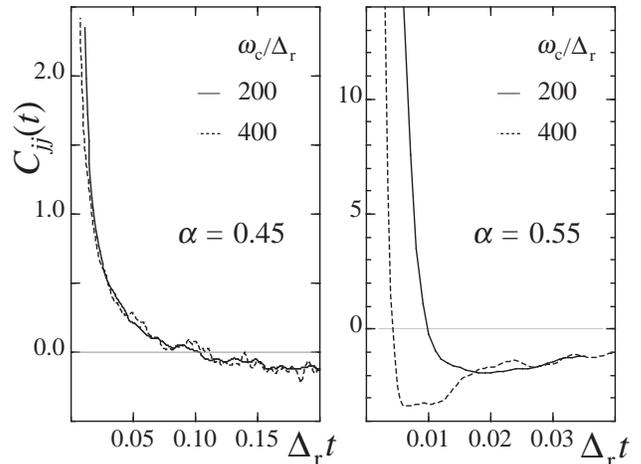}}
\caption[]{Scaling behavior of the current correlation function in the
vicinity of its first zero.\label{Fig-scale}}
\end{figure}

An expanded view of the initial decay of $C_{jj}(t)$ in the vicinity
of the critical value $\alpha_{\rm c} = 1/2$ is provided in Fig.\
\ref{Fig-scale}. In this region, the correlation time $\tau$ decreases
rapidly with increasing $\alpha$ (note the different scales on the
time axes).  Comparing results for different cutoff frequencies
$\omega_{\rm c}$, we observe markedly different scaling behavior for
$\alpha$ above or below $1/2$. For $\alpha<1/2$, $\tau$ is scaled by
$\Delta_{\rm r}$, leaving $\Delta_{\rm r}\tau$ finite; whereas for
$\alpha>1/2$, $\tau$ scales as $\omega_{\rm c}^{-1}$,
i.e. $\Delta_{\rm r}\tau$ vanishes in the scaling limit.

Critical behavior at $\alpha_{\rm c} = 1/2$ is also predicted by
perturbation theory in the bare tunneling frequency $\Delta$. In the
scaling limit, the short-time behavior of the current correlation
function is given by
\begin{equation} \label{pert}
C_{jj}(t) = {\Delta_{\rm r}^2\over 4} \cos(\pi \alpha) (\Delta_{\rm r}
t)^{-2 \alpha}.
\end{equation}
For $\alpha<1/2$, this expression (valid for $\omega_c^{-1}\ll t \ll
\Delta$) as well as the exact perturbative result (valid for $t\ll
\Delta^{-1}$) is

\noindent  positive. It follows that $\tau \gtrsim
\Delta^{-1}$, which can only be satisfied if $\tau$ scales with
$\Delta_{\rm r}$ rather than $\omega_{\rm c}$. For $\alpha>1/2$,
(\ref{pert}) is negative, and we conclude that $C_{jj}(t)$ turns
negative at a time $\tau$ of the order of $\omega_c^{-1}$.

Taken together, these results lead to a clear picture of how the
scaled dephasing time $\Delta_{\rm r}\tau$ depends on the damping
strength, shown in Fig.\ \ref{Fig-tauofalpha}. We see a gradual
decline of $\Delta_{\rm r}\tau$ to very small but finite values,
followed by a discontinuous drop at $\alpha_{\rm c} = 1/2$. The finite
value $\Delta_{\rm r}\tau\approx0.034$ at $\alpha=1/2$ emerges both
from our simulations and from
applying $C_{jj}(t) = -\ddot C(t) / 4$
to analytic results for the position correlation function
at $\alpha=1/2$\cite{Guinea,Sassetti}.

Our conclusion $\alpha_{\rm c}=1/2$ coincides exactly with the value
originally found by Chakravarty and Leggett\cite{Chakravarty-Leggett}
for the dynamics of a two-state system with a factorized initial
preparation.  In contrast, Lesage, Saleur, and Skorik\cite{Lesage} as
well as Costi and Kiefer\cite{Costi} recently found a lower value
$\alpha_{\rm c}\approx 1/3$
\noindent for the critical damping. In both their
works, the criterion of coherence was not the observation of tunneling
oscillations themselves, but the presence of inelastic peaks in
the spectral function of the position fluctuations. For {\em
strongly}\/ damped oscillations in the symmetric function $C(t)$,
however, the inelastic peaks at positive and negative frequencies
become wide enough to merge into a single peak centered at
$\omega=0$. The absence of a peak at finite frequency therefore does
{\em not}\/ necessarily indicate the absence of coherence. In order to
determine the true critical coupling, a more elaborate analysis needs
to be performed on the spectra of \cite{Lesage,Costi}.

In conclusion, we have introduced a new numerical algorithm for the
dynamics of dissipative quantum systems, which solves the dynamical
sign problem and eliminates slowing-down problems. Its validity,
efficiency and accuracy have been demonstrated for the spin-boson
model with Ohmic dissipation. We have determined the lifetime of
coherent superposition states at zero temperature and different
strengths of the damping from the current correlation function.  This
timescale decreases with increasing damping strength $\alpha$, but
remains finite (indicating quantum coherence) for all $\alpha$ below
the critical value $\alpha_{\rm c} = 1/2$. Analytic results, where
available, are reproduced with good agreement. Details of the CSQD
method and its applications to more complex problems such as the noise
spectrum of fractional quantum Hall systems will be reported
elsewhere.

This research has been supported by the National Science Foundation
under grant CHE-9528121.
CHM is a NSF Young Investigator (CHE-9257094),
a Camille and Henry Dreyfus Foundation Camille Teacher-Scholar and a
Alfred P. Sloan Foundation Fellow.
Computational resources have been
provided by the IBM Corporation under the SUR Program at USC.

\end{multicols}

\begin{references}

\bibitem{gw}
H. Grabert and H. Wipf, in {\it Advances in Solid State Physics}, 
Vol.~30, 1 (Vieweg, Braunschweig, 1990).

\bibitem{glass}
B. Golding, N.M. Zimmerman, and S.N. Coppersmith,
Phys. Rev. Lett. {\bf 68}, 998 (1992).

\bibitem{Chakravarty-Kivelson} S. Chakravarty and S. Kivelson,
Phys. Rev. Lett. {\bf 50}, 1811 (1983).

\bibitem{Chakravarty-Anderson} S. Chakravarty and P.W. Anderson,
Phys. Rev. Lett. {\bf 72}, 3859 (1994).

\bibitem{garg}
A. Garg, Phys. Rev. Lett. {\bf 77}, 964 (1996).

\bibitem{marcus}
R.A. Marcus and N. Sutin, Biochim. Biophys. Acta. {\bf 811}, 265 (1985).

\bibitem{Chakravarty-Leggett}
S. Chakravarty and A.J. Leggett,
Phys. Rev. Lett. {\bf 52}, 5 (1984).

\bibitem{Leggett}
A.J. Leggett, S. Chakravarty, A.T. Dorsey, M.P.A. Fisher, A. Garg
and W. Zwerger, Rev. Mod. Phys. {\bf 59}, 1 (1987), {\em ibid.}, {\bf
67},
%
\begin{figure}
\epsfxsize=0.75\columnwidth
\centerline{\epsffile{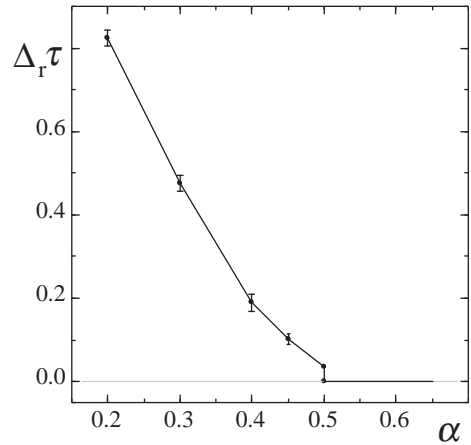}}
\caption[]{Variation of the scaled dephasing time $\Delta_{\rm r}\tau$ with
increasing damping $\alpha$.\label{Fig-tauofalpha}}
\end{figure}
\noindent 725 (1995).

\bibitem{Wipf}
H. Wipf, D. Steinbinder, K. Neumaier, P. Gutsmiedl, A. Magerl, and
A.J. Dianoux, Europhys. Lett. {\bf 4}, 1379 (1987).

\bibitem{Lesage}
F. Lesage, H. Saleur, and S. Skorik, Phys. Rev. Lett. {\bf 76}, 3388
(1996).

\bibitem{Costi}
T.A. Costi and C. Kieffer, Phys. Rev. Lett. {\bf 76}, 1683 (1996).

\bibitem{Egger}
R. Egger, H. Grabert, and U. Weiss, Phys. Rev. E {\bf 55}, R3809
(1997).

\bibitem{Strong}
S.P. Strong, Phys. Rev. E {\bf 55}, 6636 (1997).

\bibitem{Leung} K. Leung, R. Egger, and C.H. Mak, Phys. Rev. Lett. {\bf 75},
3344 (1995).

\bibitem{Chakravarty}
S. Chakravarty, Phys. Rev. Lett. {\bf 49}, 681 (1982);
A.J. Bray and M.A. Moore, Phys. Rev. Lett. {\bf 49}, 1545 (1982).

\bibitem{Guinea}
F. Guinea, Phys. Rev. B {\bf 32}, 4486, (1985).

\bibitem{alge-remark} We are not aware of systematic errors that
prevent our numerical method from reproducing this result. However, we
leave verification of this to a future publication.

\bibitem{Zurek} W.H. Zurek, Phys. Rev. D {\bf 26}, 1862 (1982).

\bibitem{Stuttgart}
M. Grifoni, M. Winterstetter, and U. Weiss, Phys. Rev. E {\bf 56}, 334
(1997).

\bibitem{WeissBuch}
U. Weiss, {\em Quantum Dissipative Systems} (World Scientific,
Singapore, 1993).

\bibitem{Feynman-Vernon}
R.P. Feynman and F.L. Vernon, Ann. Phys. (N.Y.) {\bf 24}, 118 (1963).

\bibitem{Thirumalai} D. Thirumalai, E.J. Bruskin, and B.J. Berne,
J. Chem. Phys. {\bf 79}, 5063 (1983); D. Makarov and N. Makri,
Chem. Phys. Lett. {\bf 221}, 482 (1994).

\bibitem{Cao} J. Cao, L.W. Ungar, and G.A. Voth, J. Chem. Phys. {\bf
104}, 4189 (1996).

\bibitem{cutoff-remark} A Gaussian cutoff
$\exp(-\omega^2/2\omega_{\rm c}^2)$ is advantageous here. This
modifies eq. (\ref{Delta_r}) by an insignificant numerical factor.

\bibitem{deph-remark} In the limit of zero coupling, $\tau$ will
underestimate the dephasing time.  Apart from this, none of our conclusions 
will be affected by a change in the formal definition of $\tau$.

\bibitem{Sassetti}
M. Sassetti and U. Weiss, Phys. Rev. A {\bf 41},
5383 (1990);
F. Guinea, V. Hakim and A. Muramatsu, Phys. Rev. B {\bf 32},
4410 (1985).

\end{references}
\end{document}